**Experimental and ab initio studies of the novel piperidine-containing acetylene glycols.**

**Authors:** A. Mirsakiyeva[a], D. Botkina[d,f], K. Elgammal[a], A. Ten[f], H.W. Hugosson[a], A. Delin[a,b,c], V. Yu[e,f]
a – Department of Material and Nanophysics, School of Information and Communication Technology, Electrum 229, Royal Institute of Technology (KTH), SE-16440 Kista, Sweden;
b – Department of Physics and Astronomy, Uppsala University, Box 516, 751 20 Uppsala, Sweden;
c – SeRC (Swedish e-Science Research Center), KTH, SE-10044 Stockholm, Sweden;
d – Department of Production Engineering, School of Industrial Engineering and Management, Brinellvägen 68, Royal Institute of Technology, 10044 Stockholm, Sweden;
e – Department of Chemical Engineering, Kazakh-British Technical University, Tole bi 59, 050000 Almaty, Kazakhstan;
f – Institute of Chemical Sciences, Walikhanov str. 106, 050010, Almaty city, Kazakhstan.

**Abstract**
Synthesis routes of novel piperidine-containing diacetylene are presented. The new molecules are expected to exhibit plant growth stimulation properties. In particular, the yield in a situation of drought is expected to increase. The synthesis makes use of the Favorskii reaction between cycloketones/piperidone and triple-bond containing glycols. The geometries of the obtained molecules were determined using nuclear magnetic resonance (NMR). The electronic structure and geometries of the molecules were studied theoretically using first-principles calculations based on density functional theory. The calculated geometries agree very well with the experimentally measured ones, and also allow us to determine bond lengths, angles and charge distributions inside the molecules. The stability of the OH-radicals located close to the triple bond and the piperidine/cyclohexane rings was proven by both experimental and theoretical analyses. The HOMO/LUMO analysis was done in order to characterize the electron density of the molecule. The calculations show that triple bond does not participate in intermolecular reactions which excludes the instability of novel materials as a reason for low production rate.

**1. Introduction**

Piperidine-containing compounds have shown great promise as plant growth stimulators [1-5] (Fig 1).

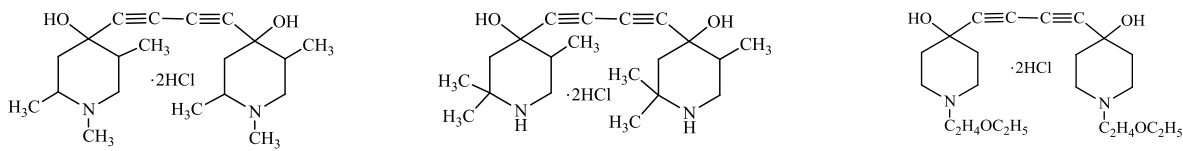

Fig 1 – Piperidine-containing diacetylene glycols

Nowadays in the world one out of eight people does not get enough food [6]. It brings the problem of hunger and malnutrition to the first position as a challenge to human health — greater than AIDS, malaria, and tuberculosis combined [7-10]. A prospective solution to the hunger problem may be offered through compounds that have the ability to increase crop yield, for example by making crops more drought resistant. It is well known that the use of piperidine-containing diacetylene glycols may increase the growth of plants multifold [1-5]. In addition, plants which were processed by piperidine-containing diacetylene glycols prove to be more resistant to drought, temperature changes and diseases [1-5]. Treatment by piperidine-containing diacetylene glycols can rapidly increase the annual harvest, especially in hot and arid regions. The synthesis of piperidine-containing diacetylene glycols is however relatively complicated. It includes ethynylation of piperidone-4 in liquid ammonia or diethyl ether in the presence of solid potassium hydroxide KOH. After extraction the purified acetylenic glycols dimerize and form diacetylene glycol. Dimerization of acetylenic glycols is provided by oxidation in the presence of cuprous chloride and pyridine [1, 2, and 5] or through hydrogenation in the presence of Raney nickel [3, 4].
In the present work a new type of compounds with potentially similar positive effects on crop yield and drought resistance was synthesized and studied both experimentally and theoretically. Piperidine-based acetylene glycols were synthesized by using so-called green chemistry methods [11] under the Favorskii reaction [12]. The Favorskii reaction can be defined as the nucleophilic attack of a terminal alkyne with acidic protons on a carbonyl group [13]. For the syntheses two alcohols with triple bond were chosen: propargyl alcohol (Fig 2.1) and 3-butyn-1-ol (Fig 2.2).

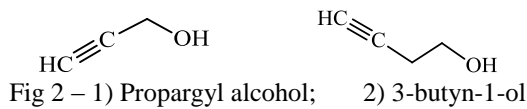

Fig 2 – 1) Propargyl alcohol;     2) 3-butyn-1-ol

Both alcohols have low crystallization temperature (-48$^o$C for propargyl alcohol and -64$^o$C for 3-butyn-1-ol) which lets the reaction take place under 0-4$^o$C without additional catalysis and formation of side gas products such as acetylene. During this study the propargyl and 3-butyn-1-ol alcohols were used in combination with 1-(2-etoxyethyl)-piperidine-4-on and cyclohexanone as model structure to create 1-(2-ethoxyethyl)-4-(3-hydroxypropynyl-1)-piperidine-4-ol, 1-(2-ethoxyethyl) - 4-(4-hydroxybutynyl-1)-piperidine-4-ol, 1-(3-hydroxypropynyl-1)-cyclohexanol and 1-(4-hydroxybutynyl-1)-cyclohexanol. However, the yields were low in the reactions under the standard conditions, whereas the duration of synthesis was long. That could be caused by the effect of the solvent, the reactivity of the raw materials with each other, or the instability of molecules and intermolecular interaction. In order to elucidate the origin of the latter possibility, we perform theoretical studies of the molecules under realistic conditions, assuming ambient temperature. We also tested the effect of van der Waals (vdW) interactions. The quantum mechanical (QM) simulations allow us to determine the detailed structure of the novel molecules (including the bond lengths, the angles and the electronic structure). Furthermore, such simulations allow us to determine whether the molecules are stable or not. This is important since if they are unstable that would be the explanation for the low production rate. In addition, using our theoretical analysis we are able to suggest possible chemical reactions to be studied further experimentally.

Numerous theoretical electronic structure studies using density functional theory (DFT) [22] and Hartree-Fock (HF) method investigating the structure and reactivity of acetylene and piperidine/cyclohexane ring containing compounds have been presented earlier [14-20]. Common to all these studies is that the triple bond has been found to have high reactivity. Also there is strong interest in studying the chemical reactivity of these novel materials being used as catalysts [14, 15]. The number of works investigates the chemical reaction mechanisms [16, 17]. DFT calculations of piperidine compounds have also been performed to determine the heat of formation [18]. An earlier investigation has convincingly demonstrated the high accuracy and agreement between DFT methods and experimental results for 4-methylpiperidine [19]. This gives confidence that these methods are suitable for further studies of this group of compounds. For 4-piperidone the ground state (HOMO) and first excited state (LUMO) were found, and the structure was determined both with theoretical (HF and DFT) and empirical (IR) methods [20]. However, we have found no previous theoretical research for molecules containing both piperidine/cyclohexane rings and triple bonds. In the theoretical part of this research, ab initio calculations of piperidine-containing acetylene glycols were performed. New molecules were built and optimized via the AVOGADRO software [21]. The electronic structure and physical properties were then studied using DFT from a fundamental point of view. Applying Car-Parrinello Molecular Dynamics [23] allowed to determine the ground-state structure of these novel compounds. The charge density difference was also studied via Quantum Espresso code [24] and visualized through XCrysDen [25].

## 2. Method

### 2.1 Experiments

Traditionally, the ethynylation reaction is carried out in an organic solvent medium (liquid ammonia, tetrahydrofuran, ether, DMSO, xylene et al.) [27-31]. Nevertheless, the use of KOH/DMSO suspension in reactions of cyclic compounds ethynylation leads to the formation of substances which do not contain an ethynyl group, making the use of 1,4-dioxane the most suitable and available solvent [32, 33]. However, the series of experiments in the medium of 1,4-dioxane was unsuccessful. The products were characterized by extremely low yields (~ 11%), as well as the presence of inseparable impurities which hinders identification. The using of 1,4-dioxane is complicated by its capability of forming explosive peroxides, and the resinification of synthesis. In addition, the crystallization temperature of 1,4-dioxane is 12 ° C, accordingly, does not allow one to carry out the reaction at lower temperatures (0-3° C).

The absence of solvent in the reaction of cyclic ketones ethynylation allows providing the reaction at low temperatures (0-5 ° C), i.e. the crystallization temperature of involved substances is below the reaction temperature. The necessity of lower temperatures is caused by the reactivity of the acetylenic alcohols and potassium hydroxide. Mixing of potassium hydroxide and propargyl alcohol occurs with the release of heat and the formation of by-products at room temperature, and decomposition to acetylene, hydrogen, formic acid and carbonic acid at a heating. Temperature decrease and the absence of the peroxide impurities in the solvent leads to increased product yields, reduction of substances resinification, and formation of fewer by-products during the reaction [31, 34].

In the general procedure the cooled (0-3$^o$C) 0.1 mol of KOH and 0.06 mol of acetylene alcohol were placed into three-necked flask, equipped with condenser, stirrer, thermometer and dropping funnel. The mixture was stirred constantly for 1 h at the constant temperature. Afterwards, 0.02 mol of cooled (4-6$^o$C) cyclic ketone was added with stirring, and maintained at temperature 0-5$^o$C for 2 h and at temperature 20-22$^o$C for 18 h. The progress of the reaction was monitored by thin layer chromatography (TLC) (eluent: chloroform/1,4-dioxane – 30:1). After finishing of the reaction, water was added to the destruction of the reaction complex. The following steps were: extraction with chloroform for separation of organic components, solvent evaporation, and re-extraction with hexane to separate the impurities.

Schemes of reactions are shown on Fig 3 and Fig 4.

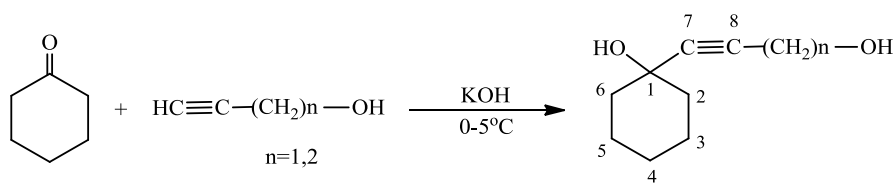

Fig 3 - Synthesis of 1-(3-hydroxypropynyl-1)-cyclohexanol and 1-(4-hydroxybutynyl-1)-cyclohexanol in solvent-free reaction medium

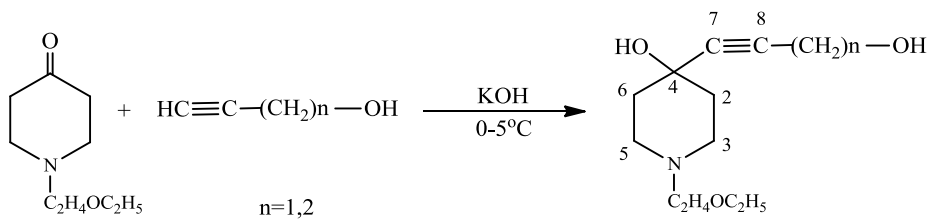

Fig 4 – Synthesis of 1-(2-ethoxyethyl) -4-(3-hydroxypropynyl-1)-piperidin-4-ol and 1-(2-ethoxyethyl) -4-(4-hydroxybutynyl-1)-piperidin-4-ol in solvent-free reaction medium

## 2.2 Theoretical calculations

The structures of the piperidine-containing acetylene glycols were built using the AVOGADRO software. QM calculations based on DFT were performed using Car-Parrinello Molecular Dynamics (CPMD). An orthorhombic simulation box with a size of $(a+8)^3$ Å, where *a* is a length of the organic molecules (varying between 20-30 Å), was used. The pseudo-potentials were of the soft norm-conserving Troullier-Martins type, with a cut-off of 90 Ry. We used the exchange-correlation functional BLYP [37, 38]. The initial quenching was combined with annealing down to $E_{kin}=0$ eV for 0.483 ps. In this context quenching refers to the converging the wavefunctions at the beginning of a run, and annealing refers to scaling of the ionic velocities towards zero. This equilibration procedure was followed by 24 ps of temperature controlled molecular dynamics calculations (velocity scaling 300 K with $\Delta T = +/-20$ K). After that the calculations were followed by 24 ps of molecular dynamics simulations at 300 K using a Nose-Hoover thermostat [39], during which we collected statistics. In order to determine the average structure and charge distribution, we extracted bond lengths and angles in every simulation step of the calculations under the Nose-Hoover thermostat. The CPMD code uses a R-ESP method to calculate the point charges from the electrostatic potential restricted to charges from the Hirshfeld method [40, 41]. For charge distribution studies an additional 4.8 ps of Nose-Hoover calculation were performed while constantly collecting charges every $10^{th}$ steps (0.00483 ps). In addition, complimentary simulations were performed including van der Waals interactions by adding the semi-empirical Grimme method [35].

Finally, charge density difference plots were generated via ground state DFT calculations using the Quantum Espresso simulation package with the setup similar to the CPMD simulations.

## 3. Results and analysis

### 3.1 Experiments

The reaction of ethynylation with cyclic ketones and acetylene alcohols in solvent-free media was characterized by satisfactory average yields, as well as the structure of product molecules were confirmed by physical and chemical methods of analysis. Thus, during the reaction in a medium without solvent, it becomes possible to examine the reaction products due to their purity and stability.

$^{13}$C NMR spectra were recorded on a Varian Mercury-300 NMR spectrometer in $CDCl_3$, HMDS as the internal standard. IR spectra were recorded on "Nicolet 5700 FT-IR" in thin film and in KBr tablets. Elemental analyses were performed on a CE-440 (EAI Exeter Analytical Inc.). A control of the reaction and a compound individuality was performed by $Al_2O_3$ thin layer chromatography (TLC) method with iodine vapors development.

The study showed that declared structures were confirmed. Corresponding peaks of ester and triple bonds, as well as hydroxyl groups were shown in IR-spectrometry data. $^{13}$C NMR spectra data also fully verify the structure of the synthesized acetylenic glycols. The most downfield signals in the range 81.16 - 89.31 ppm belong to carbon triple bond. Moreover, a carbon atom of the triple bond associated with the cycle resonates in a weaker field, because of the proximity of the hydroxyl group at $C_{1(4)}$. In propargyl derivatives there is strong shielding effect (~ 10 ppm) due to the methylene carbon adjacent to the OH group having a triple bond.

***1-(3-Hydroxypropynyl-1)cyclohexanol.*** Pale yellow oil, yield, 1.91 g (62.0%), $n^{20}_D$ 1.508. $R_f$ 0.11 (eluent: chloroform/ 1,4-dioxane – 30:1). IR (thin film): 1448.0; 1068.8 cm$^{-1}$ (O-H), 2261.6 cm$^{-1}$ (C≡C). $^{13}$C NMR δ 23.17 (C-3 and C-5), 25.13 (C-4), 39.67 (C-2 and C-6), 50.64 (CH$_2$-OH), 69.60 (C-1), 82.35 (≡**C**-CH$_2$), 89.31 (C$_1$-**C**≡). Anal. Calcd for C$_9$H$_{14}$O$_2$: C, 70.10; H, 9.15. Found: C, 70.18; H, 9.12.

***1-(4-Hydroxybutynyl-1)cyclohexanol.*** Transparent needle crystals, yield, 1, 95 g (56,9 %), mp 104,5-105$^o$C. $R_f$ 0,04 (eluent: chloroform/ 1,4-dioxane – 30:1). IR (KBr): 1450,7; 1038,6 cm$^{-1}$ (O-H), 2231,9 cm$^{-1}$ (C≡C). $^{13}$C NMR δ 23.04 ((≡C-**CH**$_2$), 23.36 (C-3 and C-5), 25.20 (C-4), 40.08 (C-2 and C-6), 61.01 (CH$_2$-OH), 68.70 (C-2 and C-6), 81.16 (≡**C**-CH$_2$), 85.96 (C$_1$-**C**≡). Anal. Calcd for C$_{10}$H$_{16}$O$_2$: C, 71.39; H, 9.59. Found: C, 71.33; H, 9.57.

***1-(2-Ethoxyethyl) -4-(3-hydroxypropynyl-1)piperidin-4-ol.*** Transparent pale-brown oil, yield, 1,23 g (46,4 %), $n^{20}_D$ 1,554. $R_f$ 0,06 (eluent: chloroform/ 1,4-dioxane – 30:1). IR (thin film): 1447,7; 1073,5 cm$^{-1}$ (O-H), 1111,0 cm$^{-1}$ (C-O-C), 2250,1 cm$^{-1}$ (C≡C). $^{13}$C NMR δ 15.11 (OCH$_2$CH$_2$OCH$_2$**C**H$_3$), 38.65 (C-3 and C-5), 50.47 (C-2 and C-6), 50.73 (CH$_2$-OH), 57.32 (O**C**H$_2$CH$_2$OCH$_2$CH$_3$), 66.43 (OCH$_2$**C**H$_2$OCH$_2$CH$_3$), 68.08 (OCH$_2$CH$_2$O**C**H$_2$CH$_3$), 73.49 (C-4), 83.74 ((≡**C**-CH$_2$), 89.01 (C$_4$-**C**≡). Anal. Calcd for C$_{12}$H$_{21}$NO$_3$: C, 63.41; H, 9.31. Found: C, 63.49; H, 9.30.

***1-(2-Ethoxyethyl) -4-(4-hydroxybutynyl-1)piperidin-4-ol.*** Transparent pale-brown oil, yield, 1,25 g (44,3 %), $n^{20}_D$ 1,593. $R_f$ 0,05 (eluent: chloroform/ 1,4-dioxane – 30:1). IR (thin film): 1445,5; 1053,7 cm$^{-1}$ (O-H), 1116,7 cm$^{-1}$ (C-O-C), 2240,7 cm$^{-1}$ (C≡C). $^{13}$C NMR δ 15.12 (OCH$_2$CH$_2$OCH$_2$**C**H$_3$), 23.02 (15-C), 39.09 (C-3 and C-5), 50.78 (C-2 and C-6), 57.45 (O**C**H$_2$CH$_2$OCH$_2$CH$_3$), 60.84 (CH$_2$-OH), 66.42 (OCH$_2$**C**H$_2$OCH$_2$CH$_3$), 69.18 (OCH$_2$CH$_2$O**C**H$_2$CH$_3$), 70.10 (C-4), 82.05 ((≡**C**-CH$_2$), 85.08 (C$_4$-**C**≡). Anal. Calcd for C$_{13}$H$_{23}$NO$_3$: C, 64.70; H, 9.61. Found: C, 64.86; H, 9.5.

### 3.2 Theoretical calculations

Below, we present our results on bond lengths and bond angles, point charge distributions, CDD, and frontier orbitals.

### 3.2.1 Structure (bond lengths and angles)

The average bond lengths for the studied molecules were determined. We here choose to concentrate our analysis on the acetylene bonds since for our purposes this is the most chemically relevant part of the electronic structure. In Table 3.2.1 we show our data for the triple bond of the substitute radical and ethers bonds. From this table, we can see that $a_2$ is significantly shorter than $a_1$ and $a_3$ (1.2 Å compare to 1.4-1.5 Å) which strongly suggests that the $a_2$ bond is a triple bond. This is an excellent agreement with the NMR results (see section 3.1). By comparing calculations with and without van der Waals corrections we conclude that the general effect of vdW corrections is minute. In addition, we note that the longer the glycol chain and the ether radical, the smaller is the difference between calculations with and without van der Waals corrections.

Table 3.2.1 - Bond lengths (in Å) from calculations using Noose-Hoover thermostat. Standard deviations are given within parentheses.

|  | a1 ($C_{ring}$-C) | a2($C_{triple}$C) | a3(C-C) | a7(C-$O_{ester}$) | a8($O_{est}$-C) |
|---|---|---|---|---|---|
| 1-(3-hydroxypropynyl-1)-cyclohexanol | 1,474 (0,030) | 1,214 (0,017) | 1,466 (0,026) | - | - |
| 1-(3-hydroxypropynyl-1)-cyclohexanol with vdW corrections | 1,476 (0,032) | 1,214 (0,015) | 1,467 (0,028) | - | - |
| 1-(4-hydroxybutynyl-1)-cyclohexanol | 1,477 (0,032) | 1,214 (0,018) | 1,468 (0,030) | - | - |
| 1-(4-hydroxybutynyl-1)-cyclohexanol with vdW corrections | 1,477 (0,032) | 1,214 (0,018) | 1,468 (0,030) | - | - |
| 1-(2-ethoxyethyl) -4-(4-hydroxybutynyl-1)-piperidine-4-ol | 1,477 (0,036) | 1,214 (0,016) | 1,469 (0,033) | 1,446 (0,031) | 1,459 (0,035) |
| 1-(2-ethoxyethyl) -4-(4-hydroxybutynyl-1)-piperidine-4-ol with vdW corrections | 1,478 (0,033) | 1,214 (0,011) | 1,468 (0,031) | 1,446 (0,029) | 1,460 (0,029) |
| 1-(2-ethoxyethyl) -4-(3-hydroxypropynyl-1)-piperidine-4-ol | 1,474 (0,029) | 1,214 (0,013) | 1,460 (0,029) | 1,448 (0,029) | 1,455 (0,030) |
| 1-(2-ethoxyethyl) -4-(3-hydroxypropynyl-1)-piperidine-4-ol with vdW corrections | 1,474 (0,029) | 1,214 (0,013) | 1,460 (0,029) | 1,448 (0,029) | 1,455 (0,030) |

To understand the structure of the novel molecules it is also essential to determine the angles between the triple bond and the OH-radical with respect to the hexane ring. From the data in the first column in Table 3.2.2 we conclude that the OH radical cannot interact with the triple bond of the alkyne radicals because of the wide angle between these two substitutes. Also one can see that the hexane ring does not deform after adding ether substitutes, neither for the hexanol, nor the piperidine rings.

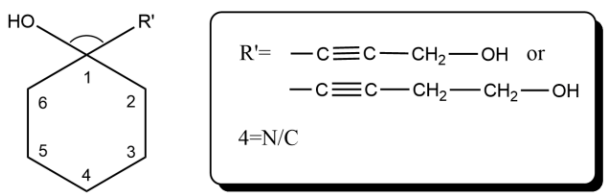

Fig 5 – Definitions of the angles in Table 3.2.2.

Table 3.2.2 – Angles (in °) from calculations using Noose-Hoover thermostat. Standard deviations are given within parentheses.

| | O-$C_{ring}$-$C_{triple}$ | $C_{ring1}$ | $C_{ring2}$ | $C_{ring3}$ | $C_{ring4}$/N | $C_{ring5}$ | $C_{ring6}$ |
|---|---|---|---|---|---|---|---|
| 1-(3-hydroxypropynyl-1)-cyclohexanol | 109,2 (3,6) | 110,4 (3,2) | 112,6 (3,2) | 111,7 (3,5) | 111,3 (3,2) | 111,8 (3,4) | 112,6 (3,2) |
| 1-(3-hydroxypropynyl-1)-cyclohexanol with vdW corrections | 108,8 (3,9) | 110,4 (3,7) | 112,6 (3,3) | 111,9 (3,5) | 111,4 (3,9) | 111,8 (3,5) | 112,5 (3,2) |
| 1-(4-hydroxybutynyl-1)-cyclohexanol | 108,6 (3,6) | 109,8 (3,0) | 112,8 (3,3) | 111,9 (3,3) | 111,7 (3,2) | 111,6 (3,4) | 112,8 (3,3) |
| 1-(4-hydroxybutynyl-1)-cyclohexanol with vdW corrections | 108,6 (3,6) | 109,8 (3,0) | 112,8 (3,3) | 111,9 (3,3) | 111,7 (3,2) | 111,6 (3,4) | 112,8 (3,3) |
| 1-(2-ethoxyethyl) -4-(4-hydroxybutynyl-1)-piperidine-4-ol | 108,7 (4,2) | 108,6 (3,5) | 112,4 (3,6) | 111,2 (3,6) | 111,5 (3,7) | 111,2 (3,3) | 112,4 (3,6) |
| 1-(2-ethoxyethyl) -4-(4-hydroxybutynyl-1)-piperidine-4-ol with vdW corrections | 109,1 (3,7) | 108,6 (3,1) | 112,5 (3,9) | 111,2 (3,5) | 111,6 (3,9) | 111,3 (3,4) | 112,5 (3,8) |
| 1-(2-ethoxyethyl) -4-(3-hydroxypropynyl-1)-piperidine-4-ol | 109,2 (3,5) | 109,5 (3,5) | 112,2 (3,3) | 111,8 (3,4) | 112,2 (4,2) | 111,3 (3,4) | 112,2 (3,4) |
| 1-(2-ethoxyethyl) -4-(3-hydroxypropynyl-1)-piperidine-4-ol with vdW corrections | 109,2 (3,5) | 109,5 (3,5) | 112,2 (3,3) | 111,1 (3,4) | 112,2 (4,2) | 111,3 (3,4) | 112,2 (3,4) |

**3.2.2 Charge distribution and charge density difference**

We also analyzed the electronic structure, and specifically the charge density, as a means to gain more insight into the chemical reactivity of the molecules. The average overall distribution of charges among all atoms in the studied molecules are shown in Fig 6a-h. However, the most revealing data is the Hirshfield point charges on the nitrogen and the oxygen atoms. As can be seen from Fig 6e-h, the nitrogen and oxygen atoms in 1-(2-ethoxyethyl) -4-(3-hydroxypropynyl-1)-piperidine-4-ol and 1-(2-ethoxyethyl) -4-(4-hydroxybutynyl-1)-piperidine-4-ol have substantial partial negative charges. This strongly suggests that these sites are relatively reactive. However, as it was shown in section 3.2.1 the nitrogen is strongly bonded to the surrounding carbon atoms and therefore cannot participate in any reactions. On the other hand, the oxygen atoms have more negative charge which can cause interaction with neighborhood molecules. Also one can see a slight effect when including the van der Waals corrections.

Our calculated point charges can be compared to those found in the General AMBER force field (GAFF) data base. For example, the sp-hybridized carbon atom is parameterized with the charge 0,360, the $sp^3$-hybridized carbon atom – 0,878, the nitrogen atom – 0,530, and the oxygen atom – 0,465. In our calculations all of molecules, both with and without van der Waals corrections, we see that two sp-hybridized carbon atoms (triple bonded atoms) inside one molecule have different charge values while in the standard force field they are equivalent. The difference in the charge can be explained by the influence of the cyclohexane/piperidine ring: carbon atoms located closer to the ring have more negative charge. On the other hand, the cyclohexane/piperidine ring also feels the influence of acetylene, hydroxy- and ether substitutes: charge values of $sp^3$-hybridized carbon atoms inside ring differs depending on position of the atom and distance between the atom and the substitutes. Also one can see the common trend for all studied structures that $sp^3$-hybridized carbon atoms closer to the triple bond have more negative charge while the same type of carbon atoms located closer to the ethers radicals contain less negative charges. However, the $sp^3$-hybridized carbon atom (atom 1 in Fig.5) connected both to the OH-group and acetylene-containing radical has strong positive charge which is approximately twice more than the GAFF value. This discrepancy can be explained by the double influence of the high electron density of the triple bond and the vicinity to the oxygen atom in the OH-group. After applying vdW corrections we found significant changes in two cases: the triple bonded carbon atoms decrease the negative charge or even change sign. Otherwise, the vdW corrections don't have a strong effect on the point charges of the studied molecules.

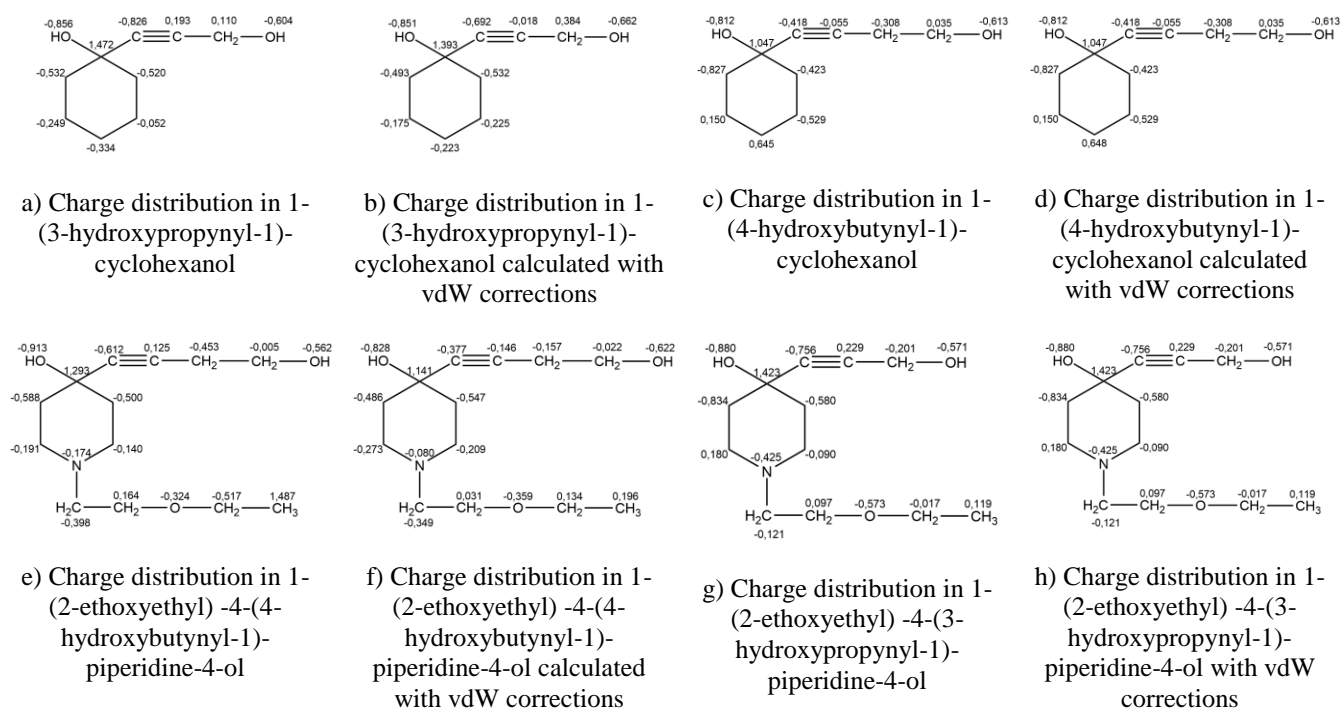

a) Charge distribution in 1-(3-hydroxypropynyl-1)-cyclohexanol

b) Charge distribution in 1-(3-hydroxypropynyl-1)-cyclohexanol calculated with vdW corrections

c) Charge distribution in 1-(4-hydroxybutynyl-1)-cyclohexanol

d) Charge distribution in 1-(4-hydroxybutynyl-1)-cyclohexanol calculated with vdW corrections

e) Charge distribution in 1-(2-ethoxyethyl) -4-(4-hydroxybutynyl-1)-piperidine-4-ol

f) Charge distribution in 1-(2-ethoxyethyl) -4-(4-hydroxybutynyl-1)-piperidine-4-ol calculated with vdW corrections

g) Charge distribution in 1-(2-ethoxyethyl) -4-(3-hydroxypropynyl-1)-piperidine-4-ol

h) Charge distribution in 1-(2-ethoxyethyl) -4-(3-hydroxypropynyl-1)-piperidine-4-ol with vdW corrections

Fig 6 – charge distributions

We also analyzed the electronic structure of the studied molecules focusing on the charge density difference, which is defined as $\Delta\rho=\rho_{mol}-\sum\rho_i$. Here, $\rho_{mol}$ and $\rho_i$ refer to the densities of the molecule and individually free atoms, respectively. Fig.7(a-h) shows isosurfaces of the charge density difference of the studied molecules where one can clearly observe regions of electronic charge accumulation (pink) and depletion (purple). The increased accumulation of electronic charge between triple bonded carbons is an indication of enhanced chemical bonding which means that the triple bond is very stable. Interestingly this is in contrast to previous studies on similar molecules [14-20]. Also it is clear that for all studied molecules a strong σ-type bonding appears between the carbon and the hydrogen atoms inside cyclohexane/piperidine ring (pink areas between ring carbons and hydrogens), while the overall ring has strong depletion of charge (purple areas around ring). This characterizes the cyclohexane/piperidine rings as nucleophile which is common for aromatic and cyclic compounds. Finally, the oxygen atom in the OH-group linked to the cyclohexane/piperidine ring attracts electron density to itself and, consequently, makes the hydrogen atom of the OH-group more positive and, therefore, highly reactive.

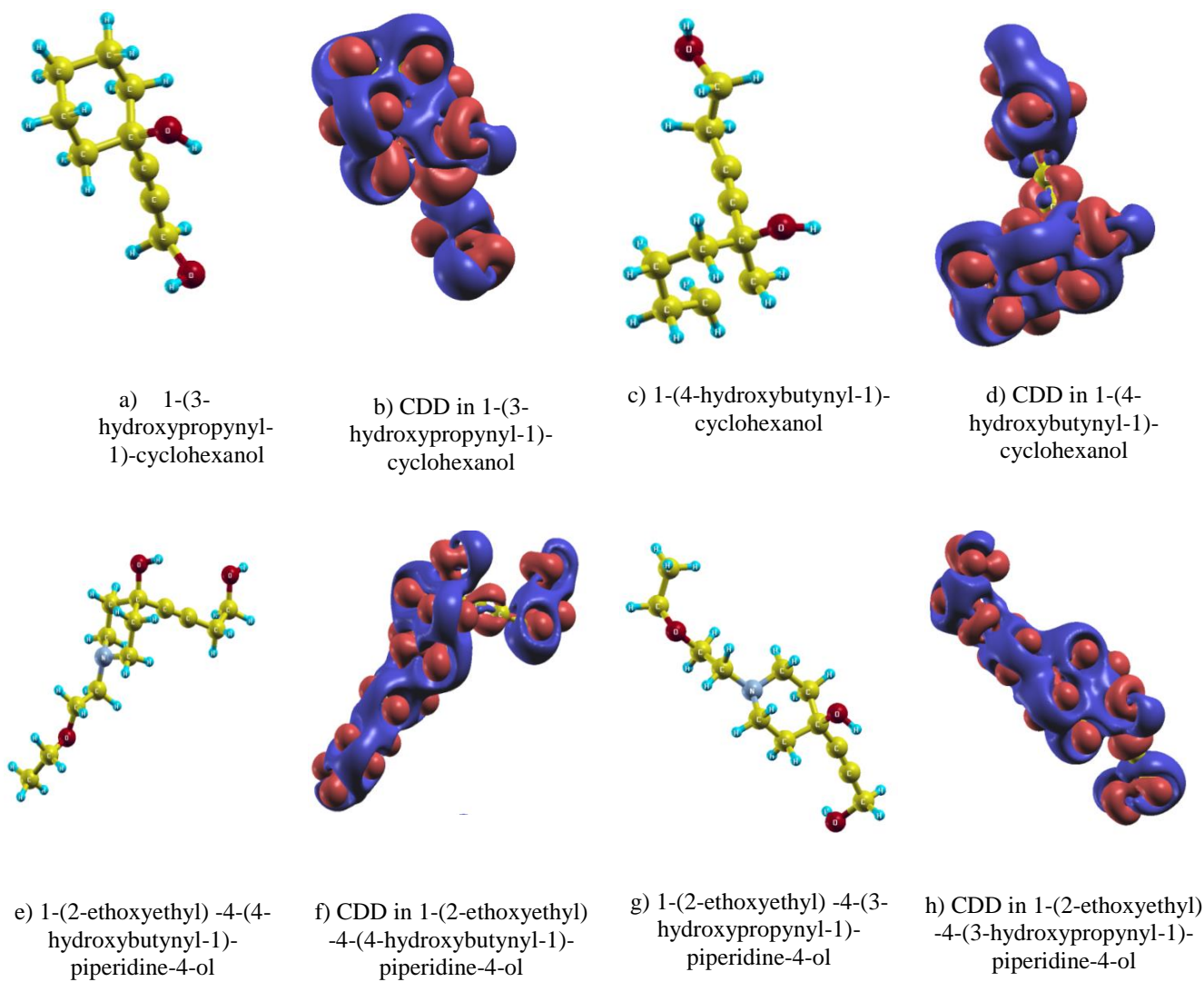

Fig 7 – Charge density difference (CDD):
Atoms: yellow – C, blue – H, red – O, light-blue – N;
Isosurfaces: pink – charge accumulation, purple – charge depletion.

In order to further analyze the nature of the bonds in the molecules we plotted the electron density along several planes (Fig. 8). We choose to present the CDD results only for the 1-(3-hydroxypropynyl-1)-cyclohexanol. The other molecules yield very similar results. In Fig.8a we show the CDD for the triple bond (the cross section on the plane is shown from side and front). One can see that the triple bonded carbon atoms (shown with green colour) have accumulation of charge. The same can be observed for the OH-group linked to the cyclohexane/piperidine ring (Fig.8b). We conclude that the triple bond is of σ-type and therefore has high stability.

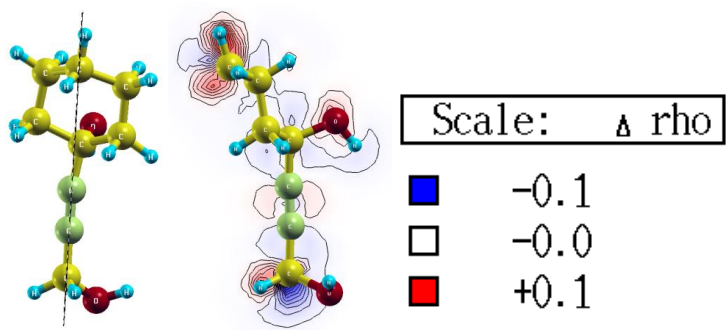

Fig 8a – The cross section of the charge density difference plot for triple bond (front and side view) of 1-(3-hydroxypropynyl-1)-cyclohexanol molecule

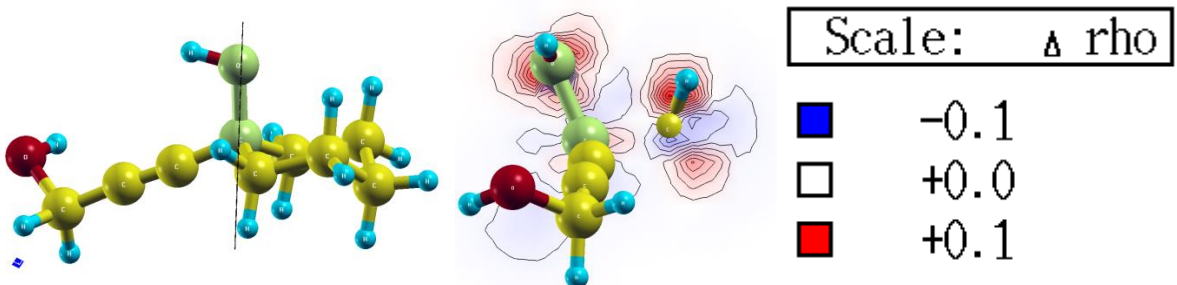

Fig 8b – The cross section of the charge density difference plot for OH-group (front and side view) of 1-(3-hydroxypropynyl-1)-cyclohexanol molecule

3.2.3 Frontier orbital analysis

The frontier orbitals are the highest occupied (HOMO) and lowest unoccupied molecular orbitals (LUMO). Again we present the frontier orbital analysis only for 1-(3-hydroxypropynyl-1)-cyclohexanol, since that analysis is representative for all studied molecules. A general result is that the HOMO and LUMO are concentrated in different parts of the molecule (Fig.9). We see that HOMO is distributed relatively equally over the molecule, whereas the LUMO is more concentrated to the acetylene substitute. The strong delocalization of the LUMO on the triple bond radical makes the triple bond available for an electrophile attack. Furthermore, it is visible that the triple bond and OH-group of glycol are interacting. The energy gap for all calculated structures is shown in Table 3.2.3. One can notice that all molecules have a high HOMO-LUMO energy difference ranging between 3.4 and 4.8 eV. The vdW interactions slightly reduces the HOMO-LUMO energy difference for the cases of 1-(3-hydroxypropynyl-1)-cyclohexanol and 1-(2-ethoxyethyl) -4-(4-hydroxybutynyl-1)-piperidine-4-ol.

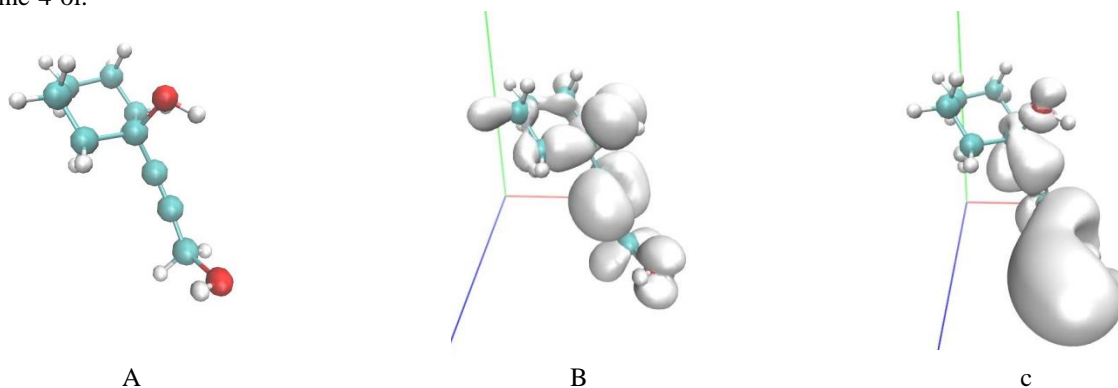

A B c

Fig 9 – a) 1-(3-hydroxypropynyl-1)-cyclohexanol structure; b) HOMO structure; c) LUMO structure;
Colors: red – oxygen, blue- carbon, white – hydrogen

Table 3.2.3 - HOMO/LUMO calculations

| | HOMO, eV | LUMO, eV | HOMO-LUMO energy gap, eV |
|---|---|---|---|
| 1-(3-hydroxypropynyl-1)-cyclohexanol | -5,9 | -1,2 | 4,7 |
| 1-(3-hydroxypropynyl-1)-cyclohexanol with vdW corrections | -5,8 | -1,1 | 4,7 |
| 1-(4-hydroxybutynyl-1)-cyclohexanol | -5,6 | -0,8 | 4,8 |
| 1-(4-hydroxybutynyl-1)-cyclohexanol with vdW corrections | -5,6 | -0,8 | 4,8 |
| 1-(2-ethoxyethyl) -4-(4-hydroxybutynyl-1)-piperidine-4-ol | -4,8 | -1.0 | 3,8 |
| 1-(2-ethoxyethyl) -4-(4-hydroxybutynyl-1)-piperidine-4-ol with vdW corrections | -4,5 | -0,9 | 3,6 |
| 1-(2-ethoxyethyl) -4-(3-hydroxypropynyl-1)-piperidine-4-ol | -4,6 | -1,2 | 3,4 |
| 1-(2-ethoxyethyl) -4-(3-hydroxypropynyl-1)-piperidine-4-ol with vdW corrections | -4,7 | -1,2 | 3,5 |

## 4. Conclusion

In this work, synthesis of new piperidine-containing acetylene glycols is reported for the first time, together with a DFT-based analysis of these compounds. Different variations of the Favorskii reaction were assessed, with the aim of improving the production rate. The results of the ethynylation of cyclohexanone and piperidone-4 in a medium without solvent at low temperature were satisfactory: the reaction proceeds with an average yield more of than 50%. The structures of the molecules correspond well to those expected. In accordance with IR-spectrometry and $^{13}$C NMR spectra, the presence of hydroxyl groups, ester and triple bonds was confirmed in the obtained molecules. Our theoretical analysis also shows that the novel molecules contain both a triple bond and a stable cyclohexane/piperidine ring. The atomic point charges were calculated using CPMD and were seen to significantly differ from the GAFF parameters. Thus, our calculated point charges may prove very useful for the parametrization of these novel molecules in case one wish to study them further using parametrized force fields. The strong influence of the OH-group, the triple-bond-containing radical and the ether-containing radical on the cyclohexane/piperidine ring was demonstrated. Also the high stability of the triple bond and high reactivity of OH-group was demonstrated through charge density difference analysis. However, our analysis of the frontier orbitals suggests possible chemical reactions such as electrophilic attack of the triple bonds which could be interesting for the further experiments. In addition, we performed several experiments with ionic liquid which gave low production. However, our theoretical calculations indicate that the low production rate is not due to structural molecular instability.